\documentstyle[12pt]{article}

\title{Quantum Corrections to
Nucleation Rates}

\author{
G. C. Marques\thanks{On leave from
Instituto de F\'{\i}sica, Universidade de S\~ao Paulo,
C.P. 20516, S\~ao Paulo, SP  01498, Brazil.}\\
{\it Texas A $\&$ M University, Dept. of Physics}\\
{\it College Station, Texas 77853, USA}
\and
Rudnei O. Ramos\\
{\it Dartmouth College, Dept. of Physics and Astronomy,}\\
{\it Hanover, NH 03755, USA}}

\date{September 1992}

\begin{document}

\maketitle

\begin{abstract}

In this paper we show how to compute in a consistent way
nucleation rates in field theory at finite temperatures, with metastable
vacuum.
Using the semiclassical approach in field theory at finite temperature we
show that the prefactor term can
be calculated explicitly (in the thin-wall approximation) and that the same
provides exponential finite temperature quantum corrections to
nucleation rates, when fluctuations around the bubble field
configuration are considered.

\vspace{0.7 cm}
\noindent
PACS number(s): 98.80.Cq, 64.60.Qb .

\end{abstract}

\section{Introduction}

The study of first-order phase transitions in cosmology has
received increased interest recently due to their possible relevance to the
physics of the early Universe. Indeed, phase transitions have been invoked
in the context of the generation of large scale
structure of
the Universe, the flatness and horizon problems$^{[1]}$ and, more
recently, the eletroweak phase transition, which could be responsible for
the generation of the cosmological baryon asymmetry$^{[2]}$.

In a first-order phase transition the Universe is believed to be in a
metastable phase (the false-vacuum) that decays to a stable phase (the
true vacuum) as the Universe expands and cools. The decay process of
the metastable phase to the stable one happens by the nucleation in
the system of droplets (or bubbles)$^{[3]}$ of the stable phase whose
dyna-\break mics will determine the completion of the phase transition$^{[4]}$.
Of fundamental relevance for the understanding of the development of the
phase transition is the determination of the bubble nucleation rate
per unit volume.

In this paper we deal with the problem of
determination, using the semiclassical approach,
of the nucleation rate in field theory at finite temperature, which
includes its definition from a droplet model field theoretical point of
view$^{[3]}$ and the evaluation of the prefactor term
appearing in the nucleation rate expression. We show that the prefactor
term    provides
an exponential finite temperature quantum correction to the
nucleation rate which has not been properly discussed in most of the studies
involving this quantity$^{[1],[5]}$. Our method of evaluation is
self-consistent and we show that the prefactor can be eva-\break luated
directly by computing the eigenvalues of the determinant terms, in the case
of a convenient bubble field configuration, or by developing an
appropriate field theoretical expansion for the determinant ratio.

The usual expression for the nucleation rate per unit volume used in the
literature is given by$^{[5]}$

\begin{equation}
\Gamma = T \left( \frac{ S_{3} (\bar{\phi},T)}{2 \pi T} \right)^{\frac{3}{2}}
\left\{ \frac{ \det [ - \nabla^{2} + V''(\phi_{f},T)]}{
\det' [ - \nabla^{2} + V''(\bar{\phi},T)]} \right\}^{\frac{1}{2}}
e^{- \frac{S_{3} (\bar{\phi},T)}{T}} \: ,
\end{equation}

\noindent
where $\bar{\phi}$ stands for the (static) bounce field configuration
describing
the bubble, $\phi_{f}$ is the false vacuum (or metastable) field configuration,
$S_{3} (\bar{\phi},T)$ is the three-dimensional Euclidean action, at
finite temperature, of the non-trivial field configuration $\bar{\phi}$,

\begin{equation}
S_{3} (\bar{\phi},T) = \int d^{3} x \left[ \frac{1}{2} ( \nabla
\bar{\phi})^{2} + V_{eff}( \bar{\phi},T) \right] \: ,
\end{equation}

\noindent
where $V_{eff}(\phi,T)$ is the effective potential at finite temperature.

What is usually done to compute the determinant ratio in (1) is the
use of dimensional analysis to approximate it by a preexponential factor
of order $T^{4}$, for example, $T_{c}^{4}$ (the critical temperature)
or $m^{4}(\phi_{f},T)$ ($m^{2}(\phi_{f},T)= \frac{d^{2} V_{eff}(\phi,T)}
{ d \phi^{2}} |_{\phi = \phi_{f}}$). This way of defining the nucleation
rate presupposes that the determinant ratio in (1) will not contribute
to the term in the exponential and that all the quantum corrections are
included, from the beginning, in the form of $S_{3}(\bar{\phi},T)$ in the
exponential.

We shall show that our way of defining the nucleation rate $\Gamma$ results
in a expression quite different of that given by (1) and
that the usual approximation for the overall preexponential factor
is not a simple function of the temperature, especially near $T_{c}$,
as observed recently by Csernai and Kapusta$^{[6]}$, that have computed
the preexponential factor in the case of a QCD phase transition.

The paper is organized in the following way: in Sec. II we review the
droplet model in field theory at finite temperature for a generic model of
a scalar field $\phi$, which could be coupled with other fields, gauge or
others scalar fields, for example. If the effective potential at finite $T$,
of the scalar field $\phi$, admits a first-order phase transition, then the
nucleation rate can be defined as being the imaginary part of the free energy,
in the approximation of a dilute gas, of droplets of the stable phase inside
the metastable phase$^{[3]}$. This imaginary free energy arises due to the
existence of a negative eigenvalue, associated with the instability of the
critical bubble.
In Sec. III we deal with the evaluation
of the prefactor appearing in the nucleation rate and show that
the same is of fundamental importance for the proper determination of the
finite temperature corrections to the exponential factor term in the
nucleation rate. In Sec. IV we study the
model of a scalar field theory coupled with fermion fields, computing the
aditional determinant factor due to the integration over the fermion fields.
In Sec. V we apply the method by choosing a simple potential which
exhibits metastability, at $T \neq 0$. Conclusions are presented in Sec. VI.

\section{Droplet Model in Field Theory}

Let us consider a system described by a scalar field $\phi$ and by a set of
fields $\chi$ (bosonic or fermionic fields), that can be coupled to $\phi$.
The partition function $Z$ of this system, at $T \neq 0$, is

\begin{equation}
Z= \int D\phi D\chi e^{-S_{Eucl}(\phi,\chi)} \: ,
\end{equation}

\noindent
where $S_{Eucl}(\phi,\chi)$ is the Euclidean action of the system
and the functional
integrations are carried over field configurations subject to periodic
(anti-periodic) conditions: $\phi(\vec{x},0)=\phi(\vec{x},\beta)$ for bosons
and $\psi(\vec{x},0)= - \psi(\vec{x},\beta)$ for fermions. $\beta$ is the
inverse of the temperature, $\beta= T^{-1}$.

If one integrates out the $\chi$ fields in (3) and makes the expansion
$\phi(\vec{x},\tau) \rightarrow \phi_{c}(\vec{x},\tau) + \eta(\vec{x},\tau)$,
where $\phi_{c}(\vec{x},\tau)$ is a field configuration that extremizes the
Euclidean action $\bar{S}_{Eucl}(\phi)$ (after integrating out the $\chi$
fields)
for the scalar field $\phi$ and $\eta(\vec{x},\tau)$ are small perturbations
around $\phi_{c}(\vec{x},\tau)$, then the effective action $\Gamma_{eff}(
\phi_{c})$ for the field configuration $\phi_{c}$, can be written as

\[
\Gamma_{eff}(\phi_{c})= \ln \int D \eta e^{-\bar{S}_{Eucl}(\phi)} \: ,
\]

\noindent
or using an expansion in terms of $\phi_{c}$ and its derivatives,

\begin{equation}
\Gamma_{eff}(\phi_{c})= \int_{0}^{\beta} d \tau \int d^{3}x \left[
- V_{eff}(\phi_{c},T) + \frac{1}{2} (\partial_{\mu}\phi_{c})^{2}
{\cal Z}(\phi_{c}) + \ldots \right] \: .
\end{equation}

\noindent
In (4), for constant field configurations, $\phi_{c}(\vec{x},\tau) =
\phi_{c}$, $V_{eff}(\phi_{c},T)$ defines the effective potential, at finite
$T$. In principle, from $V_{eff}(\phi,T)$ we can determine the order of the
phase transition. If above a certain critical temperature $T_{c}$ the system
is in a symmetric phase $\phi_{f}$ and below $T_{c}$ $V_{eff}(\phi,T)$
develops an energetically favorable phase $\phi_{t}$, then the phase
$\phi_{f}$ becomes a metastable phase and we have a first-order phase
transition in which the metastable phase $\phi_{f}$ decays to the new
favorable phase $\phi_{t}$. $T_{c}$ is defined by the condition
$V_{eff}(\phi_{f},T_{c}) = V_{eff}(\phi_{t},T_{c})$.

The process of the decay of the metastable vacuum can be seen as happening
by the nucleation of bubbles (or droplets) of the new phase $\phi_{t}$ in
the system, that in a successful phase transition will grow and coalesce
completing the phase transition.

Let us write the partition function (3), after integrating out the $\chi$
fields as

\begin{equation}
Z = \oint D \phi \exp \left\{ - \bar{S}_{Eucl}(\phi)
 \right\} \: ,
\end{equation}

\noindent
where $\bar{S}_{Eucl}(\phi)$ is an effective
Euclidean action for the scalar field $\phi$.

If the system defined by (5) admits an effective potential like that one
described above, then one can imagine the transition from the metastable
phase $\phi_{f}$ to the
stable phase $\phi_{t}$ as being preceeded by the formation of small nuclei
of the phase $\phi_{t}$ inside the phase $\phi_{f}$. Note that $\phi_{f}$
and $\phi_{t}$ are functions of the temperature, $\phi_{f(t)} \equiv
\phi_{f(t)}(T)$, given by the minima of $V_{eff}(\phi,T)$,
$\frac{d V_{eff}(\phi,T)}{d \phi}|_{\phi = \phi_{f(t)}} = 0$. Following
Langer$^{[3]}$, in a dilute gas of droplets of the phase $\phi_{t}$, one can
infer the thermodynamics of the system from the knowledge of the partition
function of a single bubble. Therefore, in a dilute gas approximation, the
principal contribution for the partition function (4) for the system can
be written as

\[
Z \simeq Z(\phi_{f}) + Z(\phi_{b})
\]

\noindent
and the partition function $Z$ of this gas of bubbles can be approximated
to$^{[3]}$

\begin{equation}
Z \simeq Z(\phi_{f}) \left[ 1 + \frac{Z(\phi_{b})}{Z(\phi_{f})} + .....
\right] \simeq Z(\phi_{f}) \exp \left[ \frac{Z(\phi_{b})}{Z(\phi_{f})} \right]
\: ,
\end{equation}

\noindent
where in the expressions above $Z(\phi_{f})$ is the partition function
for the metastable phase and $Z(\phi_{b})$ is the partition function for
the bubble in the system, or in a field theory language, $Z(\phi_{b})$ is
the partition function of the system in the presence of a bubble field
configuration $\phi_{b}$ $^{[7]}$.

For $\bar{S}_{Eucl}(\phi) = \int_{0}^{\beta} d \tau \int d^{3} x [
\frac{1}{2} (\partial_{\mu} \phi)^{2} + V(\phi)]$, the bubble field
configuration is a nontrivial solution of
the Euler-Lagrange equation:

\begin{equation}
( \frac{\partial^{2}}{\partial \tau^{2}} +
\vec{\nabla}^{2}) \phi - V'(\phi) = 0
\end{equation}

\noindent
and therefore, in the semiclassical approach, the field configuration
$\phi_{b}$
is the one that extremizes the effective action

\begin{equation}
\frac{\delta \Gamma_{eff}(\phi)}{\delta \phi}  \left|_{_{\phi=\phi_{b}}}
= 0 \right.
\: .
\end{equation}

For static field configurations, from Eq. (7), $\phi_{b}(r)$ is
a static bounce solution of the differential equation

\begin{equation}
\frac{d^{2} \phi}{d r^{2}} + \frac{2}{r} \frac{d \phi}{d r} = V'(\phi)
\end{equation}

\noindent
with boundary conditions, $\lim_{r \to +\infty} \phi = \phi_{f}$ and
$\frac{d \phi}{d r}|_{r=0} = 0$.

One can then describe $\phi_{b}(r)$ by the following spherically symmetric
field configuration

\begin{equation}
\phi_{b}(r) = \left\{
\begin{array}{ll}
\phi_{t},  & 0 < r < R- \Delta R \\
\phi_{wall},  & R - \Delta R < r < R + \Delta R \\
\phi_{f},     & r >  R + \Delta R
\end{array}
\right.
\end{equation}

\noindent
that is, $\phi_{b}(r)$ describes a bubble of radius R, of the nucleating
phase $\phi_{t}$ embedded in the metastable phase $\phi_{f}$ with the
bubble wall described by a field configuration $\phi_{wall}$, that separates
the two phases (for example, one could imagine $\phi_{wall}$ as a kink-like
field configuration). $\Delta R$ is the bubble wall thickness. The solution
(10) is a very good approximation for $\phi_{b}$ when $R >>
\Delta R$, or in the thin-wall approximation.
The simplest assumption for the validity of the thin-wall approximation is
to consider that the bubble nucleation happens at temperatures close to
$T_{c}$, that is, with relatively small supercooling$^{[5]}$. In the rest of
this work we will concentrate on this assumption, where we can approximate
the bubble field configuration quite well by $\phi_{b}(r)$ given by (10).

Let us now compute the partition function (6) using a semiclassical approach.
Within the semiclassical approach one expands the Lagrangean field $\phi$
in (6) as  $\phi \rightarrow \eta (\vec{x} ,t) + \phi_{b} (\vec{x})$  for
$Z(\phi_{b})$ and  $\phi \rightarrow \zeta(\vec{x} ,t) + \phi_{f} $  for
$Z(\phi_{f})$. $\eta(\vec{x} ,t)$ and $\zeta(\vec{x} ,t)$ are small
perturbations around the classical field configurations $\phi_{b} (\vec{x})$
and  $\phi_{f} $  respectively. Up to 1--loop order one keeps the quadratic
terms in the fields $\eta(\vec{x} ,t)$ and $\zeta(\vec{x} ,t)$ in these
expansions. In this way one can write the following expressions for
$Z(\phi_{b})$ and  $Z(\phi_{f})$, respectively,

\begin{equation}
Z(\phi_{b}) \stackrel{1-loop \: order}{\simeq} e^{-\bar{S}_{Eucl}(\phi_{b})}
\oint D \eta \exp \left\{ -\int_{0}^{\beta} d \tau \int d^{3} x \frac{1}{2}
\eta \left[ -\Box_{Eucl} + V''(\phi_{b}) \right] \eta \right\}
\end{equation}

\noindent
and

\begin{equation}
Z(\phi_{f}) \stackrel{1-loop \: order}{\simeq} e^{-\bar{S}_{Eucl}(\phi_{f})}
\oint D \zeta \exp \left\{ -\int_{0}^{\beta} d \tau \int d^{3} x \frac{1}{2}
\zeta \left[ -\Box_{Eucl} + V''(\phi_{f}) \right] \zeta \right\} \: ,
\end{equation}

\noindent
where  $\bar{S}_{Eucl}(\phi) = \int_{0}^{\beta} d \tau \int d^{3} x
[ \frac{1}{2} (\partial_{\mu} \phi)^{2} + V(\phi)]
$, $V''(\phi) = \frac{d^{2} V}{d \phi^{2}}$
and $\Box_{Eucl} = \frac{\partial^{2}}{\partial \tau^{2}} +
\vec{\nabla}^{2}$ .

Performing the functional gaussian integrals in (11) and (12) one gets the
following expression for the ratio between the partition functions,
$\frac{Z(\phi_{b})}{Z(\phi_{f})}$, appearing in (6):

\begin{equation}
\frac{Z(\phi_{b})}{Z(\phi_{f})} \stackrel{1-loop \: order}{\simeq}
\left[ \frac{ \det ( -\Box_{Eucl} + V''(\phi_{b}))_{\beta}}
{ \det ( -\Box_{Eucl} + V''(\phi_{f}))_{\beta}} \right]^{-\frac{1}{2}}
e^{-\Delta S} \: ,
\end{equation}

\noindent
where $[ \det (M)_{\beta}]^{- \frac{1}{2}} \equiv \oint D \eta \exp \left\{
- \int_{0}^{\beta} d \tau \int d^{3} x \frac{1}{2} \eta [M] \eta \right\}$
and  $\Delta S = \bar{S}_{Eucl}(\phi_{b}) - \bar{S}_{Eucl}(\phi_{f})$
is the difference
between the Euclidean actions for the field configurations
$\phi_{b}$ and $\phi_{f}$ .

The free energy of the system, ${\cal F} = - \beta^{-1} \ln Z$, up to 1-loop
approximation, from (6) and (13), can be written as

\begin{equation}
{\cal F} = - T \left[ \frac{ \det ( -\Box_{Eucl} + V''(\phi_{b}))_{\beta}}
{ \det ( -\Box_{Eucl} + V''(\phi_{f}))_{\beta}} \right]^{-\frac{1}{2}}
e^{-\Delta S} \: .
\end{equation}

It is well known that the determinant for the bubble field configuration in
(14) has a negative eigenvalue,
that signals the presence of a metastable state, and that it has also three
zero eigenvalues related with the translational invariance of the bubble
in the three dimensional space. Because of the negative eigenvalue, the free
energy ${\cal F}$ is then imaginary. However the imaginary of
${\cal F}$ can be related exactly with the
nucleation rate of bubbles of the phase $\phi_{t}$ (the stable
vacuum) inside the metastable phase $\phi_{f}$, as shown by Langer$^{[3]}$.
At finite temperatures, Affleck, ref. [8], showed that the nucleation
rate $\Gamma$ is given by

\begin{equation}
\Gamma \equiv \frac{| \omega_{-} |}{\pi} \frac{Im {\cal F}}{\beta^{-1}} \: ,
\end{equation}

\noindent
where $| \omega_{-} |$ is the frequency of the unstable mode.

\section{Evaluation of the Determinants}

Let us now compute the ratio of the determinants appearing in
the nucleation rate, Eq. (14), and show that the same provides
a finite temperature correction to the exponential term $\Delta S$ .

One remembers that $\Delta S$ is given by\footnote{We are supposing that
$\bar{{\cal L}}_{Eucl}(\phi)$ denotes an ``effective Euclidean action'' for
the field $\phi$, where possible other fields coupled with $\phi$ has
been integrated out. In this way $V(\phi)$ includes not only the
classical potential for $\phi$ but also can include corrections ($T \neq 0$)
coming from the integration of that fields.}

\begin{equation}
\Delta S = \beta \int d^{3} x \left[\bar{ {\cal L}}_{Eucl}(\phi_{b}) -
\bar{{\cal L}}_{Eucl}(\phi_{f}) \right] = \frac{\Delta E}{T} \: .
\end{equation}

\noindent
If one
uses the bubble field configuration $\phi_{b}(r)$ as given by (10), then in
the thin-wall approximation one obtains the following expression for
$\Delta E$ in (16):

\begin{equation}
\Delta E = - \frac{4 \pi R^{3}}{3} \Delta V + 4 \pi R^{2} \sigma_{0} \: ,
\end{equation}

\noindent
where $\Delta V = V(\phi_{f}) - V(\phi_{t})$ is the potential difference
between the false and true vacua (the metastable and stable vacua,
respectively) and $\sigma_{0}$ is the surface tension of
the bubble wall (with no corrections due fluctuations around the bubble
wall field configuration $\phi_{wall}$)

\begin{equation}
\sigma_{0} \simeq \int_{-\Delta R}^{+ \Delta R} d r \left[
\bar{{\cal L}}_{Eucl}(\phi_{wall}) -\bar{ {\cal L}}_{Eucl}(\phi_{f})
\right] \: .
\end{equation}

\noindent
Remind that in the above equations and in $\phi_{b}(r)$, given by (10),
$\phi_{f(t)} \equiv \phi_{f(t)}(T)$, are the minima of the finite
temperature effective potential, that is, $\phi_{f(t)}$ are given by

\begin{equation}
\frac{d V_{eff}(\phi,T)}{d \phi}|_{\phi = \phi_{f(t)}} = 0 \: .
\end{equation}

{}From (17) one can see that $\Delta E$ can be associated with
the activation energy
of one bubble of radius R. What we are going to show is that
the determinant ratio in (14) will give a finite temperature correction
to this bubble activation energy, where this correction comes exactly from
fluctuations around the bubble field configuration $\phi_{b}$.

The computation of the ratio of determinants in (14) can be done by using two
approaches. The first one involves obtaining directly the eigenvalues
of the determinants in Eq. (14). The second one consists in developing
a consistent expansion for the determinant ratio.

\subsection{Evaluation of the prefactor in terms of the eigenvalues}

The evaluation of the determinants in (14) can be done by
computing directly, if possible, the eigenvalues of the differential equations

\begin{equation}
\left[ - \Box_{Eucl} + V''(\phi_{f}) \right] \varphi_{f}(i) =
\varepsilon_{f}^{2}(i) \varphi_{f}(i)
\end{equation}

\noindent
and

\begin{equation}
\left[ - \Box_{Eucl} + V''(\phi_{b}) \right] \varphi_{b}(i) =
\varepsilon_{b}^{2}(i) \varphi_{b}(i) \: .
\end{equation}

\noindent
In momentum space one writes, $\varepsilon^{2} = \omega_{n}^{2} + E^{2}$,
where $\omega_{n} = \frac{2 \pi n}{\beta} \:,\:\:n = 0,\pm 1,\pm 2, ...$,
for bosons (for fermion fields $\omega_{n} = \frac{(2 n + 1) \pi}{\beta}$).
{}From (20) and (21) one can write the determinant ratio in (14) as

\begin{eqnarray}
\left[ \frac{ \det ( -\Box_{Eucl} + V''(\phi_{f}))_{\beta}}
{ \det ( -\Box_{Eucl} + V''(\phi_{b}))_{\beta}} \right]^{\frac{1}{2}}
&=& \exp \left\{ \frac{1}{2} \ln \left[ \frac{ \det ( -\Box_{Eucl} +
V''(\phi_{f}))_{\beta}}
{ \det ( -\Box_{Eucl} + V''(\phi_{b}))_{\beta}} \right] \right\} =
\nonumber \\
&=& \exp \left\{ \frac{1}{2} \ln \left[ \frac{ \prod_{n=- \infty}^{+ \infty}
\prod_{i} \left( \omega_{n}^{2} + E_{f}^{2}(i) \right) }{
\prod_{n= - \infty}^{+ \infty} \prod_{j} \left( \omega_{n}^{2} + E_{b}^{2}(j)
\right) } \right] \right\} \: .
\end{eqnarray}

Using the identity:

\begin{equation}
\prod_{n = 1}^{+ \infty} \left( 1 + \frac{z^{2}}{n^{2}} \right) =
\frac{\sinh (\pi z)}{ \pi z}
\end{equation}

\noindent
and taking into account that we have in (22) a negative and
three zero eigenvalues, one obtains for (22) the expression:

\begin{eqnarray}
\left[ \frac{ \det ( -\Box_{Eucl} + V''(\phi_{f}))_{\beta}}
{ \det ( -\Box_{Eucl} + V''(\phi_{b}))_{\beta}} \right]^{\frac{1}{2}}
&=& \frac{T^{4}}{i |E_{-}|} \frac{\beta \frac{|E_{-}|}{2}}{
\sin \left( \beta \frac{|E_{-}|}{2} \right)} \left[ \frac{\Delta S}{2 \pi}
\right]^{\frac{3}{2}}  \nonumber \\
& \times & \exp  \left\{ \sum_{i} \left[ \frac{\beta}{2} E_{f}(i) +
\ln \left( 1 - e^{- \beta E_{f}(i)} \right) \right]  \right. \nonumber \\
&-& \left.
\sum_{j} \; ' \left[ \frac{\beta}{2} E_{b}(j) + \ln \left( 1 -
e^{- \beta E_{b}(j)} \right) \right] \right\} \: .
\end{eqnarray}

\noindent
The factor $\left[ \frac{\Delta S}{2 \pi}\right]^{\frac{3}{2}} $
in the right hand side of (24) comes from the contribution of
the zero eigenvalues$^{[9]}$. The prime in $\sum_{j}$ is a reminder that
we have excluded the negative and the three zero eigenvalues from
the sum. $E_{-}^{2}$ denotes the negative eigenvalue ($|\omega_{-}| =
|E_{-}|$, in (15) ).

Substituting (24) in (14) and using the bubble field configuration
$\phi_{b}(r)$, given by (10), one obtains the following expression for
the nucleation rate $\Gamma$, given by (15):

\begin{equation}
\Gamma = {\cal A} T^{4}
\exp \left[ - \frac{ \Delta F(T)}{T} \right] \: ,
\end{equation}

\noindent
where we have denoted by ${\cal A}$ the adimensional factor
$\alpha \left[ \frac{\Delta E}{2 \pi T} \right]^{\frac{3}{2}}$ with
$\alpha$ given by

\begin{equation}
\alpha = \frac{1}{\pi} \frac{ \frac{|E_{-}|}{2 T}}{ \sin \left( \frac{
|E_{-}|}{2 T} \right)}
\end{equation}

\noindent
and $\Delta E$ is given by (17). $\Delta F(T)$ in (25) is given by

\begin{equation}
\Delta F(T) = - \frac{4 \pi R^{3}}{3} \Delta U(T) + 4 \pi R^{2} \sigma (T) \:,
\end{equation}

\noindent
where

\begin{eqnarray}
\Delta U(T) &=& V(\phi_{f}) - V(\phi_{t}) + T \int \frac{d^{3} k}{(2 \pi)^{3}}
\ln \left[ 1 - e^{- \beta \sqrt{ \vec{k}^{2} + m^{2}(\phi_{f})}} \right]
- \nonumber \\
&-& T \int \frac{d^{3} k}{(2 \pi)^{3}} \ln \left[ 1 - e^{- \beta \sqrt{
\vec{k}^{2} + m^{2}(\phi_{t})}} \right]
\end{eqnarray}

\noindent
and

\begin{eqnarray}
\sigma (T) &=& \sigma_{0} + \frac{T}{ 4 \pi R^{2}} \left\{ \sum_{j} \; '
\left[ \frac{ \beta}{2} E_{wall}(j) + \ln \left( 1 - e^{- \beta E_{wall}(j)}
\right) \right] - \right. \nonumber \\
&-& \left. \sum_{i} \left[ \frac{ \beta}{2} E_{f}(i) + \ln \left( 1 -
e^{- \beta
E_{f}(i)} \right) \right] \right\} \: .
\end{eqnarray}

\noindent
In (27) we have used again the thin-wall approximation. In (28) we have
substituted the discrete sums by integrals over momenta (for the constant
field configurations $\phi_{f}$ and $\phi_{t}$, we have the continuum
eigenvalues, $E_{f}^{2} = \vec{k}^{2} + m^{2}(\phi_{f})$ and
$E_{t}^{2} = \vec{k}^{2} + m^{2}(\phi_{t})$, respectively, with
$m^{2}(\phi_{f}) = \frac{d^{2} V(\phi)}{d \phi^{2}} |_{\phi = \phi_{f}}$ and
$m^{2}(\phi_{t}) = \frac{d^{2} V(\phi)}{d \phi^{2}} |_{\phi = \phi_{t}}$ ).
The terms, like $\int d^{3} k \sqrt{ \vec{k}^{2} + m^{2}
(\phi)}$, that should appear in (28) are ultraviolet divergent and can be
subtracted from the theory by the introduction of the usuals counterterms of
renormalization that render the theory finite$^{[10]}$.

In (29) $E_{wall}(j)$ are the eigenvalues related with the bubble wall
field configuration $\phi_{wall}$. The problem then is reduced to the
computation of these eigenvalues for a field configuration describing the
bubble wall.

It is easy to see that $\Delta U(T)$ as given by (28) is exactly the finite
temperature effective potential difference between the false and true
vacua, as expected$^{[11]}$. The second term in the right hand side in (29)
clearly represents the finite temperature contribution for the surface
tension $\sigma_{0}$, coming from the 1-loop finite temperature quantum
corrections due to fluctuations around the bubble wall field configuration
$\phi_{wall}$.

\subsection{A field theoretical expansion for the determinants}

The second approach for the computation of the determinants
in (14) consists in developing a simple field theoretical expansion for it.
Let us write for the ratio of the determinants the following expression

\begin{eqnarray}
\left[ \frac{ \det ( -\Box_{Eucl} + V''(\phi_{f}))_{\beta}}
{ \det' ( -\Box_{Eucl} + V''(\phi_{b}))_{\beta}} \right]^{\frac{1}{2}}
&=& \exp \left\{ \frac{1}{2} Tr \; \ln \left[-\Box_{Eucl} +
V''(\phi_{f}) \right]_{\beta} - \right. \nonumber \\
&-& \left. \frac{1}{2} Tr' \; \ln \left[-\Box_{Eucl} +
V''(\phi_{b}) \right]_{\beta} \right\} \: ,
\end{eqnarray}

\noindent
where we have used in (30) the identity $ \ln \det \hat{M} = Tr \; \ln
\hat{M}$ and the prime in both sides denote that the negative and the zero
modes have been omitted.

Formally one can write (30) as

\begin{equation}
\left[ \frac{ \det ( -\Box_{Eucl} + V''(\phi_{f}))_{\beta}}
{ \det' ( -\Box_{Eucl} + V''(\phi_{b}))_{\beta}} \right]^{\frac{1}{2}}
= \exp \left\{ - \frac{1}{2} Tr \; \ln \Bigl [ 1 + G_{\beta}(\phi_{f})
\left[ V''(\phi_{b}) - V''(\phi_{f}) \right] \Bigr ] \right\} \: ,
\end{equation}

\noindent
where

\begin{equation}
G_{\beta}(\phi_{f}) = \frac{1}{ - \Box_{Eucl} + m^{2}(\phi_{f}) }
\end{equation}

\noindent
is just the free propagator, at finite temperature, for the scalar field
$\phi$, with mass squared given by
$m^{2}(\phi_{f}) = V''(\phi_{f})$.

If we expand the natural logarithm in (31) in powers of $G_{\beta}(\phi_{f})
\left[ V''(\phi_{b}) - V''(\phi_{f}) \right]$, we get formally

\[
Tr \; \ln \left\{ 1 + G_{\beta}(\phi_{f})
\left[ V''(\phi_{b}) - V''(\phi_{f}) \right] \right\} =
\]

\vspace{0.7 cm}

\begin{equation}
= \hspace{11.0 cm} + \: \: \ldots \: ,
\end{equation}

\vspace{1.0 cm}

\noindent
where the dashed lines correspond to the background-like field
$\left[ V''(\phi_{b}) - V''(\phi_{f}) \right]$ and the internal lines denote
the propagator $G_{\beta}(\phi_{f})$. The expression (33) can be written
as

\begin{eqnarray}
Tr \; \ln \left\{ 1 + G_{\beta}(\phi_{f})
\left[ V''(\phi_{b}) - V''(\phi_{f}) \right] \right\} &=& \sum_{m=1}^{+ \infty}
\frac{ (-1)^{m+1} }{m} \int d^{3} x  \left[V''(\phi_{b}) - V''(\phi_{f})
\right]^{m}  \times \nonumber \\
&\times &  \sum_{n= -\infty}^{+ \infty} \int \frac{ d^{3} k}
{(2 \pi)^{3}} \frac{1}{ \left[ \omega_{n}^{2} + \vec{k}^{2} + m^{2}(\phi_{f})
\right]^{m} } \: .
\end{eqnarray}

The sum in $m$ in (34) can be formally performed and one obtains

\begin{equation}
Tr \; \ln \left\{ 1 + G_{\beta}(\phi_{f})
\left[ V''(\phi_{b}) - V''(\phi_{f}) \right] \right\} =
\int d^{3} x  \sum_{n= -\infty}^{+ \infty} \int \frac{ d^{3} k}
{(2 \pi)^{3}} \ln \left[ 1 + \frac{ V''(\phi_{b}) - V''(\phi_{f}) }
{\omega_{n}^{2} + \vec{k}^{2} + m^{2}(\phi_{f}) } \right] \: .
\end{equation}

\noindent
Substituting (35) in (31) and again taking into account that we have
eliminated the negative and zero modes from the determinant in (14),
which
provides us with that factor multipling the exponential in (24),
one obtains an
expression for the nucleation rate $\Gamma$ like
(25), but now with $\Delta F(T)$ given by (using (10) as a solution for
$\phi_{b}$ and the thin-wall approximation)

\begin{equation}
\Delta F(T) = - \frac{ 4 \pi R^{3}}{3} \Delta V_{eff}(T) + 4 \pi R^{2}
\sigma (T) \: ,
\end{equation}

\noindent
where

\begin{equation}
\Delta V_{eff} (T) = V(\phi_{f}) - V(\phi_{t}) - \frac{1}{2 \beta}
\sum_{n= - \infty}^{+ \infty} \int \frac{ d^{3} k}{(2 \pi)^{3}} \ln \left[ 1 +
\frac{ m^{2}(\phi_{t}) - m^{2}(\phi_{f}) }
{\omega_{n}^{2} + \vec{k}^{2} + m^{2}(\phi_{f}) } \right]
\end{equation}

\noindent
and

\begin{equation}
\sigma (T) = \sigma_{0} + \frac{1}{4 \pi R^{2}} \int d^{3} x \frac{1}{2 \beta}
\sum_{n= - \infty}^{+ \infty} \int \frac{ d^{3} k}{(2 \pi)^{3}} \ln \left[ 1 +
\frac{V''(\phi_{wall}) - V''(\phi_{f}) }
{\omega_{n}^{2} + \vec{k}^{2} + m^{2}(\phi_{f}) } \right] \: .
\end{equation}

Expression (37) can easily be identified with $\Delta U(T)$, given by
(28), if one performs the sum in $n$ in (37) by using the identity

\begin{eqnarray}
\sum_{n= -\infty}^{+ \infty} \ln \left[ \frac{ \omega_{n}^{2} +
E_{t}^{2}(\vec{k}) }{ \omega_{n}^{2} + E_{f}^{2}(\vec{k}) } \right] &=&
\beta E_{t}(\vec{k}) + 2 \ln \left( 1 - e^{- \beta E_{t}(\vec{k}) } \right)
- \nonumber \\
&-& \beta E_{f}(\vec{k}) - 2 \ln \left( 1 - e^{- \beta E_{f}(\vec{k}) } \right)
\end{eqnarray}

\noindent
and as it was done in (28), eliminating the ultraviolet divergent terms,
we obtain

\begin{equation}
\Delta V_{eff}^{Ren} (T) = V_{eff}^{Ren} (\phi_{f},T) - V_{eff}^{Ren}
(\phi_{t},T) \: ,
\end{equation}

\noindent
where

\begin{equation}
V_{eff}^{Ren}(\phi,T) = V(\phi) + T \int \frac{d^{3} k}{(2 \pi)^{3}}
\ln \left( 1 - e^{- \beta \sqrt{\vec{k}^{2} + m^{2}(\phi)}} \right)
\end{equation}

\noindent
is the renormalized effective potential at finite temperature.

Expression (38) for $\sigma (T)$ can be easily identified as being the
surface tension at finite temperature for the bubble wall. If one writes
(38) as

\begin{equation}
\sigma (T) = \frac{1}{4 \pi R^{2} \beta} \left[ \Gamma_{eff} (\phi_{wall},T) -
\Gamma_{eff} (\phi_{f},T) \right] \: ,
\end{equation}

\noindent
where $\Gamma_{eff} (\phi,T)$ is given by

\begin{equation}
\Gamma_{eff} (\phi,T) = \int_{0}^{\beta} d \tau \int d^{3} x \left\{
\bar{{\cal L}}_{Eucl}(\phi) + \frac{1}{2 \beta} \sum_{n= -\infty}^{+ \infty}
\int \frac{d^{3} k}{(2 \pi)^{3}} \ln \left[ \omega_{n}^{2} +
\vec{k}^{2} + m^{2}(\phi) \right] \right\} \: ,
\end{equation}

\noindent
$\Gamma_{eff}(\phi,T)$, as given by (43), is the effective action at finite
temperature for a field configuration $\phi$ at 1-loop order$^{[12]}$.
$\sigma (T)$, given by (42), is the definition of the
surface tension in field theory$^{[12]}$ .

It is easily shown$^{[7]}$ that in a high temperature limit ($\beta
\rightarrow 0,\: T \rightarrow \infty$) the terms in the expansion (33) that
have higher superficial degree of divergence will contribute with a leading
power in $T$. This contribution is just the first graph in (33)
($m = 1$ in (34))
such that at high temperatures one can consider only the following simple
term of the graphic expansion in (33):

\begin{equation}
\hspace{3.0 cm} = \int d^{3} x \left[ V''(\phi_{b}) - V''(\phi_{f}) \right]
\sum_{n = - \infty}^{+ \infty} \int \frac{d^{3} k}{(2 \pi)^{3}}
\frac{1}{ \omega_{n}^{2} + \vec{k}^{2} + m^{2}(\phi_{f})} \: ,
\end{equation}

\noindent
making the computation of $\Delta F(T)$ and then of $\Gamma$ very simple
with a bubble field confi-\break guration $\phi_{b}$ as given by (10).

\section{A Scalar Field $\phi$ Coupled to Fermion Fields}

Let us now study a model of a
scalar field $\phi$ coupled with massless fermion fields
$\psi$,
with Lagrangian density given by

\begin{equation}
{\cal L} = \frac{1}{2} (\partial_{\mu} \phi)^{2} - V(\phi) - g \phi
\bar{\psi} \psi + i \bar{\psi} \not{\! \partial} \psi \: .
\end{equation}

\noindent
$V(\phi)$ in (45) is such that the finite temperature effective potential
of the scalar field $\phi$ admits a first-order phase transition. For example
with $V(\phi)$ given by Eq. (67), Sec. V.

The partition functions $Z(\phi_{b})$ and $Z(\phi_{f})$ in (6) are now
defined by the field expansions  $\phi(\vec{x},t) \rightarrow
\phi_{b}(\vec{x}) + \eta(\vec{x},t)$ and $\psi(\vec{x},t) \rightarrow
\psi(\vec{x},t)$ for $Z(\phi_{b})$ and $\phi(\vec{x},t) \rightarrow
\phi_{f} + \zeta (\vec{x},t)$ and $\psi(\vec{x},t) \rightarrow
\psi(\vec{x},t)$ for $Z(\phi_{f})$. Again, $\phi_{f(t)} \equiv
\phi_{f(t)}(T)$ are determined by the minima of the effective potential
$V_{eff}(\phi,T)$. Up to 1-loop order one gets the following
expressions for $Z(\phi_{b})$ and $Z(\phi_{f})$, respectively,

\begin{eqnarray}
Z(\phi_{b}) &=& e^{- S_{Eucl}(\phi_{b})} \oint D \eta \exp \left\{
- \int_{0}^{\beta} d \tau \int d^{3} x \frac{1}{2} \eta \left[ - \Box_{Eucl} +
V''(\phi_{b}) \right] \eta \right\} \times \nonumber \\
&\times & \oint D \bar{\psi} D \psi \exp \left\{ - \int_{0}^{\beta}
d \tau \int d^{3} x \bar{\psi} \left[ - \not{\! \partial} - i g \phi_{b}
\right] \psi \right\}
\end{eqnarray}

\noindent
and

\begin{eqnarray}
Z(\phi_{f}) &=& e^{- S_{Eucl}(\phi_{f})} \oint D \zeta \exp \left\{
- \int_{0}^{\beta} d \tau \int d^{3} x \frac{1}{2} \zeta \left[ - \Box_{Eucl} +
V''(\phi_{f}) \right] \zeta \right\} \times \nonumber \\
&\times & \oint D \bar{\psi} D \psi \exp \left\{ - \int_{0}^{\beta}
d \tau \int d^{3} x \bar{\psi} \left[ - \not{\! \partial} - i g \phi_{f}
\right] \psi \right\} \: ,
\end{eqnarray}

\noindent
where the functional integration over the bosonic (fermionic) fields satisfy
periodic (antiperiodic) boundary conditions in Euclidean time.

As in (14) the free energy of the system can be written now as

\begin{equation}
{\cal F} = - T \left[ \frac{ \det \left( - \Box_{Eucl} + V''(\phi_{f})
\right)_{\beta} }{ \det \left( - \Box_{Eucl} + V''(\phi_{b}) \right)_{\beta}}
\right]^{\frac{1}{2}} \left[ \frac{ \det ( - \not{\! \partial} - i g
\phi_{b})_{\beta} }{ \det ( - \not{\! \partial} - i g \phi_{f} )_{\beta} }
\right] e^{- \Delta S} \: ,
\end{equation}

\noindent
where $\Delta S = S_{Eucl}(\phi_{b}) - S_{Eucl}(\phi_{f})$, with, from (45),
$S_{Eucl}(\phi) = \int_{0}^{\beta} \int d^{3} x \left[ \frac{1}{2}
(\partial_{\mu} \phi)^{2} + V(\phi) \right]$.

The determinant ratio of the bosonic part in (48) can be computed in exactly
same way as before and we get the following expression for the nucleation
rate $\Gamma$ for the model (45),

\begin{equation}
\Gamma = {\cal A} T^{4}
\left[ \frac{ \det ( - \not{\! \partial} - i g
\phi_{b})_{\beta} }{ \det ( - \not{\! \partial} - i g \phi_{f} )_{\beta} }
\right] \exp \left[ - \frac{ \Delta F(T)}{T} \right] \: ,
\end{equation}

\noindent
where ${\cal A}$ and $\Delta F(T)$ have analogous identification with
the terms appearing in (25). $\Delta F(T)$ can be given by both expressions
(27) or (36) obtained by the eigenvalue method or by the graphic expansion,
respectively.

The fermionic determinant ratio appearing in (49) can be calculated by the
same way as the bosonic one. We shall see that it provides also a correction
to the exponential term of the nucleation rate $\Gamma$, that is, a
fermionic correction to the factor $\Delta F(T)$ in eq. (49).

If one uses the identity (which follows from charge-conjugation invariance):

\begin{eqnarray}
\left[ \det ( - \not{\! \partial} - i g \phi) \right]^{2} &=&
\det ( - \not{\! \partial} -i g \phi).det( - \not{\! \partial} + i g \phi)
= \nonumber \\
&=& \det \left[ ( -\Box_{Eucl} + g^{2} \phi^{2} ) 1_{4 \times 4} -
i g \gamma_{Eucl}^{\mu} \partial_{\mu} \phi \right] \: ,
\end{eqnarray}

\noindent
where $1_{4 \times 4}$ is the $4 \times 4$ unit matrix. From (50) the
denominator of the fermionic determinant ratio in (49) can be written as
($\phi_{f}$ is the metastable vacuum field configuration)

\begin{equation}
\det ( - \not{ \! \partial} - i g \phi_{f})_{\beta} =
\left[ \det ( - \Box_{Eucl} + g^{2} \phi_{f}^{2} )_{\beta} \right]^{
\frac{1}{2}} \: .
\end{equation}

\noindent
For the determinant involving the bubble field configuration one can again use
a configuration like (10) for $\phi_{b}$, which describes a radially symmetric
bubble solution. Making the following choice for the Dirac matrix in the
radial direction

\begin{equation}
\gamma_{r} = i \left(
\begin{array}{cc}
1_{2 \times 2} & 0 \\
0 & - 1_{2 \times 2}
\end{array}
\right) \: ,
\end{equation}

\noindent
where $1_{2 \times 2}$ denotes a $ 2 \times 2$ unit matrix, one can write
$\det ( - \not{\! \partial} - ig \phi_{b}(r) )_{\beta}$ as

\begin{equation}
\det ( - \not{\! \partial} - ig \phi_{b}(r) )_{\beta} =
\det \hat{ \Omega}^{(+)}(\phi_{b}) . \det \hat{\Omega}^{(-)}(\phi_{b}) \: ,
\end{equation}

\noindent
where

\begin{equation}
\hat{\Omega}^{(\pm)}(\phi_{b}) = - \Box_{Eucl} + g^{2} \phi_{b}^{2} \pm
g \frac{\partial \phi_{b}}{\partial r} \: .
\end{equation}

The fermionic determinant ratio in (49) can be written then as (using again
that $\ln \det \hat{M} = Tr \ln \hat{M}$ )

\begin{eqnarray}
\frac{ \det ( - \not{\! \partial} - i g \phi_{b} )_{\beta} }{ \det (
- \not{\! \partial} - i g \phi_{f} )_{\beta} } &=&
\exp \left\{ Tr \ln ( - \not{\! \partial} - i g \phi_{b} )_{\beta} -
Tr \ln ( - \not{\! \partial} - i g \phi_{f} )_{\beta} \right\}
= \nonumber \\
&=& \exp \left\{ Tr \ln \left[ -\Box_{Eucl} + g^{2} \phi_{b}^{2} +
g \frac{ \partial \phi_{b}}{\partial r} \right]_{\beta} +
Tr \ln \left[ - \Box_{Eucl} + g^{2} \phi_{b}^{2} - g \frac{ \partial
\phi_{b}}{\partial r} \right]_{\beta} - \right. \nonumber \\
&-& \left. 2 Tr \ln \left[ - \Box_{Eucl} + g^{2} \phi_{f}^{2} \right]_{\beta}
\right\} \: .
\end{eqnarray}

As in (31) one can write (55) in the form

\begin{eqnarray}
\frac{ \det ( - \not{\! \partial} - i g \phi_{b} )_{\beta} }{ \det (
- \not{\! \partial} - i g \phi_{f} )_{\beta} } &=&
\exp \left\{ Tr \ln \left[ 1 + S_{\beta}(\phi_{f}) \left[ g^{2} ( \phi_{b}^{2}
- \phi_{f}^{2} ) + g \frac{ \partial \phi_{b}}{\partial r} \right] \right]
+ \right. \nonumber \\
&+& \left. Tr \ln \left[ 1 + S_{\beta}(\phi_{f}) \left[ g^{2} ( \phi_{b}^{2}
- \phi_{f}^{2} ) - g \frac{ \partial \phi_{b}}{\partial r} \right] \right]
\right\} \: ,
\end{eqnarray}

\noindent
where

\begin{equation}
S_{\beta}(\phi_{f}) = \frac{1}{ -\Box_{Eucl} + m_{F}^{2}(\phi_{f}) }
\end{equation}

\noindent
is the fermionic analogous of $G_{\beta}(\phi_{f})$ given by (32) and
$m_{F}^{2}(\phi_{f}) = g^{2} \phi_{f}^{2}$. The expression in the exponent
in (56) can be written as a series expansion like (34):

\begin{eqnarray}
Tr \ln \left[ 1 + S_{\beta}(\phi_{f}) \left[ g^{2} ( \phi_{b}^{2}
- \phi_{f}^{2} ) \pm g \frac{ \partial \phi_{b}}{\partial r} \right] \right]
&=& \sum_{m= 1}^{+ \infty} \frac{(-1)^{m+1}}{m} \int d^{3}x \left[ g^{2}
( \phi_{b}^{2}
- \phi_{f}^{2} ) \pm g \frac{ \partial \phi_{b}}{\partial r} \right]^{m}
\times \nonumber \\
&\times & \sum_{n= -\infty}^{+ \infty} \int
\frac{d^{3} k}{ (2 \pi)^{3}} \frac{1}{ \left[ \bar{\omega}_{n}^{2} +
\vec{k}^{2}
+ m_{F}^{2}(\phi_{f}) \right]^{m} } \: ,
\end{eqnarray}

\noindent
where $\bar{\omega}_{n} = \frac{(2 n + 1) \pi}{\beta}$, for fermionic fields.
As before, (58) can be expressed as a graphic expansion similar to (33)
with the
propagators $G_{\beta}(\phi_{f})$ exchanged now by $S_{\beta}(\phi_{f})$
and the external lines given by $g^{2} ( \phi_{b}^{2}
- \phi_{f}^{2} ) + g \frac{ \partial \phi_{b}}{\partial r}$ or
$ g^{2} ( \phi_{b}^{2}
- \phi_{f}^{2} ) - g \frac{ \partial \phi_{b}}{\partial r}$ .

When (56) is substituted in (49) and using the bubble field configuration
$\phi_{b}$ as given by (10), in the thin-wall approximation one obtains

\begin{equation}
\Gamma = {\cal A} T^{4}
\exp \left[ - \frac{ \Delta F^{B+F}(T)}{T} \right] \: ,
\end{equation}

\noindent
where $\Delta F^{B+F}(T)$ is given by

\begin{equation}
\Delta F^{B+F}(T) = - \frac{4 \pi R^{3}}{3} \Delta V_{eff}^{B+F}(T) +
4 \pi R^{2} \sigma^{B+F}(T) \: ,
\end{equation}

\noindent
where $\Delta V_{eff}^{B+F}(T) = V_{eff}^{B+F}( \phi_{f},T) -
V_{eff}^{B+F}( \phi_{t},T)$ is the effective potential difference between
the false and true vacua for the model (45). Up to 1-loop order one obtains,

\begin{eqnarray}
V_{eff}^{B+F} (\phi ,T) &=& V(\phi) + T \int \frac{d^{3} k}{ (2 \pi)^{3}}
\ln \left( 1 - e^{ - \beta \sqrt{\vec{k}^{2} + m_{B}^{2}(\phi) }} \right)
- \nonumber \\
&-& 4 T \int \frac{d^{3} k}{ (2 \pi)^{3}} \ln \left( 1 +  e^{ - \beta
\sqrt{\vec{k}^{2} + m_{F}^{2}(\phi) }} \right) \: ,
\end{eqnarray}

\noindent
where $m_{B}^{2}(\phi) = V''(\phi)$ and $m_{F}^{2}(\phi) =
g^{2} \phi^{2}$ are the boson and fermion effective masses, respectively,
in the background field $\phi$. $\sigma^{B+F}(T)$ in (60) is given by

\begin{equation}
\sigma^{B+F} (T) = \frac{1}{4 \pi R^{2} \beta} \left[ \Gamma_{eff}^{B+F}
(\phi_{wall},T) -
\Gamma_{eff}^{B+F} (\phi_{f},T) \right]
\end{equation}

\noindent
with

\begin{eqnarray}
\Gamma_{eff}^{B+F} (\phi,T) &=& \int_{0}^{\beta} d \tau \int d^{3} x
\left\{ \frac{1}{2} (\partial_{\mu} \phi)^{2} + V(\phi) + \frac{1}{2 \beta}
\sum_{n=-\infty}^{+ \infty}
\int \frac{d^{3} k}{
(2 \pi)^{3}} \ln \left[ \omega_{n}^{2} + \vec{k}^{2} + m_{B}^{2}(\phi)
\right]  \right. \nonumber \\
&-& \left. \frac{2}{\beta} \sum_{n=-\infty}^{+ \infty}
\int \frac{d^{3} k}{(2 \pi)^{3}} \ln \left[ \bar{\omega}_{n}^{2} + \vec{k}^{2}
+ m_{F}^{2}(\phi) + g | \vec{\nabla} \phi | \right] \right\} \: .
\end{eqnarray}

\noindent
$\Gamma_{eff}^{B+F}(\phi,T)$ is the effective action, up to 1-loop order,
for the model (45). As in (42), (62) defines the surface tension for the
bubble wall.

The determinat ratio in (55) can also be derived by the eigenvalue method
if one knows how to compute the eigenvalues of the differential equation
for the field configuration $\phi_{wall}$ describing the bubble wall,

\begin{equation}
\left[ - \Box_{Eucl} + g^{2} \phi_{wall}^{2} \pm g \frac{\partial \phi_{wall}}
{\partial r} \right] \varphi_{wall}^{(\pm)} (j) = [ \varepsilon_{wall}^{(\pm)}
(j) ]^{2} \varphi_{wall}^{(\pm)} (j) \: ,
\end{equation}

\noindent
where, in momentum space, $[ \varepsilon_{wall}^{(\pm)}(j) ]^{2} =
\bar{\omega}_{n}^{2} + [ E_{wall}^{(\pm)}(j)]^{2}$. For the vacuum
fields $\phi_{f}$ and $\phi_{t}$ we have the continuum eigenvalues (in
momentum space)

\[
[\varepsilon_{f}^{F}(\vec{k})]^{2} = \bar{\omega}_{n}^{2} +
[E_{f}^{F}(\vec{k})]^{2}
\:, \:\:\: [E_{f}^{F}(\vec{k})]^{2} = \vec{k}^{2} + g^{2} \phi_{f}^{2}
\]
\begin{equation}
\end{equation}
\[
[\varepsilon_{t}^{F}(\vec{k})]^{2} = \bar{\omega}_{n}^{2} +
[E_{t}^{F}(\vec{k})]^{2}
\:, \:\:\: [E_{t}^{F}(\vec{k})]^{2} = \vec{k}^{2} + g^{2} \phi_{t}^{2}
\]

\noindent
for the false and true vacua, respectively.

Using these eigenvalues in (55) one obtains an expression for (49)
like (59), with
$\Delta F^{B+F}(T)$ given again by (60) and as it was shown in the bosonic
case, from (65), the expression for $\Delta V_{eff}^{B+F}(T)$ remains the same
as that one given in (60). The expression for $\sigma^{B+F}(T)$ is now
rewritten in terms of the eigenvalues for the bubble wall field configuration
$\phi_{wall}$, such that, from (64) and taking into account the bosonic part
of $\sigma^{B+F}(T)$ given by (29), one can write

\begin{eqnarray}
\sigma^{B+F}(T) &=& \sigma^{B}(T) - \frac{1}{4 \pi R^{2} \beta} \left\{
\sum_{j_{(+)},j_{(-)}} \left[ \beta E_{wall}^{(\pm)}(j_{(\pm)}) +
2 \ln \left( 1 + e^{-\beta E_{wall}^{(\pm)}(j_{(\pm)})} \right) \right]
- \right. \nonumber \\
&-& \left. \sum_{i} \left[ 2 \beta E_{f}^{F}(i) + 4 \ln \left( 1 +
e^{- \beta E_{f}^{F}(i)} \right) \right] \right\} \: ,
\end{eqnarray}

\noindent
where $\sigma^{B}(T)$ is given by (29) and $E_{f}^{F}(i) = \sqrt{ \vec{k}^{2}
+ m_{F}^{2}(\phi_{f})} $.

\section{Simple Example}

Let us now illustrate our method
for a simple model that exhibits metastability and for which we
can estimate the eigenvalues related with a bubble wall field
configuration $\phi_{wall}$.

Consider the model of a scalar field $\phi$ coupled to fermion fields
$\psi$ with Lagrangian density given by (45) and potential $V(\phi)$
given by

\begin{equation}
V(\phi ,h) = - \frac{m^{2}}{2} \phi^{2} + \frac{\lambda}{4 !} \phi^{4}
- h \phi \: ,
\end{equation}

\noindent
where $h$ is assumed to be a constant external current.

It is well known that the potential (67) exhibits metastability when one
varies the sign of $h$$^{[3],[13]}$. $V(\phi ,h)$ has two minima,
$\phi_{f}(h)$ and $\phi_{t}(h)$. For $h>0$  $\phi_{t}$ describes the stable
phase and $\phi_{f}$ the metastable phase. For $h<0$  the rules of
$\phi_{f}(h)$ and $\phi_{t}(h)$ are reversed. At finite temperatures,
we can always find values for the Yukawa coupling $g$ and for the
constant $h$, such that the condition of existence of metastability
is satisfied$^{[14]}$, with $\phi_{f(t)}(h,T)$ describing the metastable
(stable) phases, respectively. In the limit $h \rightarrow
0^{\pm}$, the vacua solutions $\phi_{f,t}(h,T)$ tend to the limit $ \pm
\phi_{0}(T) = \pm \left( \frac{6 m^{2}(T)}{\lambda} \right)^{\frac{1}{2}}$,
where $m^{2}(T)$ is the finite temperature effective mass for the scalar
field $\phi$ (in the high-temperature limit, $m^{2}(T) \simeq
m^{2} - \frac{\lambda T^{2}}{24} - \frac{g T^{2}}{6}$. ) and
one of the non-trivial static solutions of the Euler-Lagrange equation for the
field $\phi$ in model (45), with potential given by (67), is

\begin{equation}
\phi_{kink}(x_{L}) = \phi_{0}(T) \tanh \left( \frac{m(T)}{\sqrt{2}} x_{L}
\right) \: ,
\end{equation}

\noindent
which describes a planar interface between the two vacua $\pm \phi_{0}(T)
\:\: (h=0)$. $x_{L}$ is the longitudinal component of the spacial
vector ( $\vec{x} = ( x_{L},\vec{x}_{T} )$ ).

For suficiently small values of $h$, the bubble field configuration
$\phi_{b}(r)$ is well approximated by the radially symmetric solution$^{[13]}$:

\begin{equation}
\phi_{b}(r) \simeq \frac{1}{2} ( \phi_{f} + \phi_{t}) + \frac{1}{2}
( \phi_{f} - \phi_{t}) \tanh \left[ \frac{m(T)}{\sqrt{2}} (r - R) \right] \: ,
\end{equation}

\noindent
which describes a bubble of radius $R$ of the nucleating vacuum $\phi_{t}
\equiv \phi_{t}(T)$
(the true vacuum) embedded in the metastable vacuum $\phi_{f} \equiv
\phi_{f}(T)$ (the false
vacuum). From (69), the thin-wall approximation is equivalent to considering
$ R>> m^{-1}(T)$, where $\Delta R$, the bubble wall thickness, is proportional
to $m^{-1}(T)$.

{}From (29) and (66) we see that we have to compute the eigenvalues of the
differential equations (approximating for small values of $h$)

\[
\left[ - \vec{\nabla}^{2} + \frac{\lambda}{6} \left( 3 \phi_{wall}^{2}
- \frac{6 m^{2}}{\lambda} \right) \right] \varphi_{wall}^{B} = E_{wall}^{2}
\varphi_{wall}^{B}
\]
\begin{equation}
\end{equation}
\[
\left[ - \vec{\nabla}^{2} + g^{2} \phi_{wall}^{2} \pm g | \vec{\nabla}
\phi_{wall} | \right] \varphi_{wall}^{(\pm)} = \left[ E_{wall}^{(\pm)}
\right]^{2} \varphi_{wall}^{(\pm)}
\]

\noindent
for the bosonic and fermionic terms, respectively.

If one writes $\varphi_{wall} = \Psi_{n,l}(r) \Phi_{l,m}(\theta,\varphi)$ ,
and as $\phi_{wall}$ can be written as a radially symmetric solution,
(70) is given by

\[
\left[ - \frac{d^{2}}{d r^{2}} - \frac{2}{r} \frac{d}{dr} +
\frac{l(l+1)}{r^{2}} +
\frac{\lambda}
{6} \left( 3 \phi_{wall}^{2} - \frac{6 m^{2}}{\lambda} \right) \right]
\Psi_{n,l}(r) =
E_{n,l}^{2} \Psi_{n,l}(r)
\]
\begin{equation}
\end{equation}
\[
\left[ - \frac{d^{2}}{d r^{2}} - \frac{2}{r} \frac{d}{dr} + \frac{l(l+1)}
{r^{2}} + g^{2} \phi_{wall}^{2} \pm g \frac{d \phi_{wall}}{dr} \right]
\Psi_{n,l}^{(\pm)}(r) = \left[ E_{n,l}^{(\pm)} \right]^{2} \Psi_{n,l}^{(\pm)}
(r)
\]

\noindent
for the bosonic and fermionic terms, respectively.

For small enough $h$ one can consider the bubble wall field configuration
$\phi_{wall}$ as given basically by a field configuration like (68), that is,
a kink-like field configuration and (71) is reduced to

\[
\left[ - \frac{d^{2}}{d r^{2}} - \frac{2}{r} \frac{d}{dr} +
\frac{l(l+1)}{r^{2}} -
m^{2} + 3 m^{2}(T) - 3 m^{2}(T) \mbox{sech}^{2} \left( \frac{m(T)}{\sqrt{2}}
r \right)
\right] \Psi_{n,l}(r) =
E_{n,l}^{2} \Psi_{n,l}(r)
\]
\begin{equation}
\end{equation}
\[
\left[ - \frac{d^{2}}{d r^{2}} - \frac{2}{r} \frac{d}{dr} + \frac{l(l+1)}
{r^{2}} + \frac{m^{2}(T)}{2} \left[ S^{2} - S(S \pm 1) \mbox{sech}^{2}
\left(\frac{m(T)}{\sqrt{2}} r \right) \right] \right]
\Psi_{n,l}^{(\pm)}(r) = \left[ E_{n,l}^{(\pm)} \right]^{2} \Psi_{n,l}^{(\pm)}
(r) \: .
\]

\noindent
In (72)  $S^{2} = \frac{12 g^{2}}{\lambda}$.
As $g$ and
$\lambda$ are arbitrary constants in the model, $S$ is an arbitrary positive
constant. Choosing $S$ as a positive integer number, the eigenvalues of (72),
for low temperatures,
can be easily estimated (the differential equations in (72) are equivalent
to the
Schr\"odinger equation for a Posch-Teller potential). The eigenvalues
$E_{n,l}^{2}$ and $\left[ E_{n,l}^{(\pm)} \right]^{2}$ are given by (see ref.
[15] for instance)

\begin{equation}
E_{n,l}^{2} = \left\{
\begin{array}{ll}
\frac{l(l+1) - 2}{R^{2}},  & n=0 \\
\frac{l(l+1)}{R^{2}} + \frac{3}{2} m^{2} + {\cal O}(\frac{T^{2}}{m^{2}
R^{2}})
, & n= 1 \\
k^{2} + 2 m^{2} + \frac{l(l+1)}{R^{2}} +{\cal O}(\frac{T^{2}}{m^{2} R^{2}})
,  & n \rightarrow k  \:\:\mbox{(continuum)}
\end{array}
\right.
\end{equation}

\noindent
and

\begin{equation}
\left[ E_{n,l}^{(\pm)} \right]^{2} = \left\{
\begin{array}{l}
\frac{l(l+1) - 2}{R^{2}} + m^{2} p(S- \frac{p}{2}) +
{\cal O}(\frac{T^{2}}{m^{2} R^{2}}) \:, \:\: p= \left\{
\begin{array}{ll}
0,1,2,...,S-1 \: ,& \mbox{for (+)} \\
1,2,...,S-1 \: ,& \mbox{for (--)}
\end{array}
\right. \\
k^{2} + \frac{l(l+1)}{R^{2}} + \frac{1}{2} m^{2} S^{2} +
{\cal O}(\frac{T^{2}}{m^{2} R^{2}})
\end{array}
\right. \: ,
\end{equation}

\noindent
where, in the equations above, $R$ is the bubble radius.

For the bosonic eigenvalues we have $\varepsilon_{bos}^{2} = \left( \frac{
2 \pi j}{\beta} \right)^{2} + E_{n,l}^{2}$ (at $T \neq 0$). Therefore
, for $j=0$, $E_{0,0}^{2}$ corresponds to a negative mode, which is associated
with the instability of the critical bubble.
For $j=0$, $E_{0,1}^{2}$ ($m=0,\pm 1$)
represents the three translational modes related to the space translation
invariance of the bubble.
At high temperatures we still may associate the negative and the zero
eigenvalues with $\frac{ l(l+1)-2}{R^{2}}$, for $l=0$ and $l=1$, respectively.
Note that, at $T=T_{c}$, the critical radius of the bubble must go to
infinity, $R_{cr}(T_{c}) \rightarrow \infty$, and therefore the negative
eigenvalue $E_{0,0}^{2} = E_{-}^{2} = - \frac{2}{R^{2}}$ vanishes, as expected.

Taking the limit $h \rightarrow 0^{\pm}$  is equivalent to taking the limit
of infinite bubble radius, $R \rightarrow \infty$. In this case the bubble
wall solution $\phi_{wall}$ tends to the planar wall interface solution
$\phi_{kink}$, eq. (68), and the eigenvalues (73) and (74) become that
of the planar wall solution. (In (73) and (74), in the limit $h \rightarrow
0^{\pm}$
($R \rightarrow \infty$), the relevant contribution of the $l$--dependence
comes from large values of $l$  such that the sum over $l$, of the eigenvalues,
$\sum_{l=0}^{+ \infty} ( 2 l + 1)$, can be replaced by a continuum integration
in the two-dimensional momenta parallel to the wall's surface, $\int d^{2}
k_{\parallel}$. )

The evaluation of expression (62), for the surface tension, in terms of
the eigenvalues, of course, is not an easily task. However, as shown in the
last two sections, we still can use the expansion for the determinants, Eqs.
(34) and (58). For example, in the limit $h \rightarrow 0^{\pm}$, we can
easily estimate the extra contribution to the surface tension, coming
from fluctuations around the wall field configuration $\phi_{wall}$.
{}From (34) and (58) one gets, in leading order in $T$ (that is, from the
tadpole graphs, like the one in (44) ), that these extra contribution for
the surface tension $
\sigma_{0}$, given by (18), is given by (approximating $\phi_{wall}(r)$
by (68), $h \rightarrow 0^{\pm}$)

\begin{equation}
\sigma_{T} = \sigma_{T}^{B} + \sigma_{T}^{F} \: ,
\end{equation}

\noindent
where

\begin{equation}
\sigma_{T}^{B} = \frac{1}{2} \int_{- \infty}^{+ \infty} d x_{L}
\left[ V''(\phi_{kink}(x_{L}) ) - V''(\phi_{0}) \right]
\int \frac{d^{3}k}{(2 \pi)^{3}} \frac{1}{ \sqrt{ \vec{k}^{2} + m_{B}^{2}
(\phi_{0})} \left( e^{\beta \sqrt{ \vec{k}^{2} + m_{B}^{2}(\phi_{0})}}
- 1 \right) }
\end{equation}

\noindent
for the bosonic contribution and

\begin{equation}
\sigma_{T}^{F} = 2 \int_{- \infty}^{+ \infty} d x_{L}
\left[ g^{2} \phi_{kink}^{2}(x_{L}) - g^{2} \phi_{0}^{2} \right]
\int \frac{d^{3}k}{(2 \pi)^{3}} \frac{1}{ \sqrt{ \vec{k}^{2} + m_{F}^{2}
(\phi_{0})} \left( e^{\beta \sqrt{ \vec{k}^{2} + m_{F}^{2}(\phi_{0})}}
+ 1 \right) }
\end{equation}

\noindent
is the fermionic contribution. In (76) and (77), $\phi_{kink}(x_{L})$
is given by (68), $\phi_{0}^{2} = \frac{6 m^{2}(T)}{\lambda}$, $m_{B}^{2}
(\phi_{0}) = \frac{d^{2} V(\phi)}{d \phi^{2}} |_{\phi= \phi_{0}(T)}$
and  $m_{F}^{2}(\phi_{0}) = g^{2} \phi_{0}^{2}(T) = \frac{6 m^{2}(T) g^{2}}
{\lambda}$.

In the high-temperature limit, $T >> m$ ($\lambda \sim g^{2}$, $ \lambda
<< 1$), we obtain that

\[
\sigma_{T}^{B} \simeq - \frac{ \sqrt{2} m(T) T^{2}}{4} + {\cal O} (T) \: ,
\]
\begin{equation}
\end{equation}
\[
\sigma_{T}^{F} \simeq - \frac{ \sqrt{2} m(T) g^{2} T^{2}}{\lambda} +
{\cal O} (T)
\: .
\]

\noindent
Using that $S^{2} = \frac{12 g^{2}}{\lambda}$, we obtain

\begin{equation}
\sigma_{T} \simeq - \frac{ \sqrt{2} m(T) T^{2}}{4} \left[ 1 + \frac{S^{2}}{3}
\right] + {\cal O} (T) \: .
\end{equation}

\noindent
where $m^{2}(T)$, in leading order in $T$, is given by $m^{2}(T) \simeq
m^{2} - \frac{ \lambda T^{2}}{24} - \frac{g T^{2}}{6}$.

\section{Conclusions}

In this paper we have shown how to compute
the nucleation rate in field theory at finite temperature
and also we have shown how the
prefactor term (the ratio of the determinant) contribute to the exponential
term in the classical nucleation rate, showing that it provides a
finite temperature quantum correction to the exponential term.

The procedure we have used is completely consistent and it is based
on the usual description of the decay of metastable states
developed first by Langer$^{[3]}$, translated here for finite
temperature field theories.
Actually, the final expressions that we have obtained
for the nucleation rate, Eq. (25) or Eq. (59), are similar the one
proposed by Langer, in which the droplet energy $\Delta E$ is replaced
by the droplet free energy $\Delta F(T)$, when one takes
into account the prefactor term.

For the evaluation of the prefactor term (the determinant ratio) we have
used two methods. The first method is based on the explicit evaluation
of the eigenvalues. The
second procedure is based on a field theoretic expansion for the
determinant ratio, such that we avoid the task of computing the eigenvalues of
the differential equations. Both methods depend on the knowledge, at least
in an approximate way, of a field configuration describing the bubble wall.

We have shown (using the thin-wall approximation) that the nucleation rate
has a form that is much like the ``classical'' one (without taking into
account the determinants). In the general expression for the nucleation rate
$\Gamma$, we have shown that $\Delta F(T)$ is given by

\[
\Delta F(T) = - \frac{4 \pi}{3} R^{3} \Delta V_{eff}(T) + 4 \pi R^{2}
\sigma(T) \: ,
\]

\noindent
where $\Delta V_{eff}(T)$ is just the effective potential contribution
that contains the ``classical'' result plus a quantum correction that is
calculable in field theory. The surface term $\sigma(T)$ is a sum of
two contributions: a ``classical'' term plus a temperature dependent quantum
correction. The general structure of $\sigma(T)$ is given by Eq. (66).

Due to the similarity between $\Delta F(T)$ and the free energy of the
bubble, there is apparently no distintion between our results
and analogous ones presented in the literature$^{[1],[5]}$. This will be
true as long as one computes $\Delta V_{eff}(T)$ and $\sigma(T)$ in a
consistent way. We have given in this paper examples of how to carry on
such a calculation consistently.

It is interesting to contrast the expression that we get for the nucleation
rate with  expression (1). We have shown that all quantum corrections, due to
fluctuations around the bubble solution,
to $\Delta S$ in the exponential term in (14) are in the determinant ratio.
For the nucleation rate associated to a critical bubble (when $\Delta F(T)$
assumes its maximum value), from (25) and (26) we get the following
complete expression for $\Gamma$:

\begin{equation}
\Gamma = I_{0}(R_{cr}) e^{-\frac{16 \pi \sigma^{3}(T)}{3T
\Delta V_{eff}^{2}(T)}}
\end{equation}

\noindent
with the preexponential term $I_{0}(R_{cr})$ given by

\begin{equation}
I_{0}(R_{cr}) = \frac{1}{\pi} \frac{ \frac{1}{\sqrt{2} T R_{cr}(T)}}
{\sin \left[ \frac{1}{\sqrt{2} T R_{cr}(T)} \right] }
\left[ \frac{ \Delta E (R_{cr})}{2 \pi T} \right]^{\frac{3}{2}} T^{4} \: .
\end{equation}

\noindent
In (80) we have used (from the expression for $\Delta F(T)$) that

\[
R_{cr}(T) = \frac{2 \sigma (T)}{\Delta V_{eff}(T)}
\]

\noindent
and in (81) we have used, from Eq. (73), that the negative eigenvalue is
given by \break $E_{-}^{2} = - \frac{2}{R_{cr}(T)^{2}}$, for the model
(45) with
potential given by (67).

We remark also that in
ref. [6] the preexponential factor $I_{0}(R)$ for critical bubbles was
estimated, in the case that $\Gamma$ represents the probability per unit
volume per unit time to nucleate a critical bubble of the hadronic
phase inside the quark phase, to be$^{[6]}$

\begin{equation}
I_{0}(R_{cr}) = \frac{4}{\pi} \left( \frac{ \sigma}{3T} \right)^{\frac{3}{2}}
\frac{ \sigma (\zeta_{q} + \frac{4 \eta_{q}}{3}) R_{cr}}{\xi_{q}^{4}
(\Delta \omega)^{2}} \: ,
\end{equation}

\noindent
where $\zeta_{q}$ and $\eta_{q}$ are viscosity coefficients, $\xi_{q}$ is
the correlation length in the quark-gluon phase, $\Delta \omega$ is the
difference of enthalpy density between the quark-gluon phase and
$\sigma$ is the surface tension, assumed there to be temperature independent.
The authors in [6] have pointed out then that
the usual approximation for the
preexponential term, like $T^{4}$ or $T_{c}^{4}$, can not be
a good approximation,
since, from (82), the preexponential is a complicate function of the
temperature, diverging as $T \rightarrow T_{c}$. The same is true here, where
the preexponential factor is given by Eq. (81).

Despite the fact that we have carried out our analysis by using a thin-wall
approximation, that is only valid for nucleation happening at temperatures
close to $T_{c}$, or equivalently, for small supercooling, it is possible
to extent the method for other cases of interest, for example, thick-wall
bubbles$^{[5,16]}$, provided that we can find an approximate solution
$\phi_{b}(r)$, describing the bubble field configuration.

\vspace{1.0cm}

\begin{center}

{\large \bf Acknowledgements}
\end{center}

\vspace{0.7 cm}

We thank M.Gleiser for useful discussions and for a critical reading of the
manuscript and A. Linde for discussions. We also thank the kind
hospitality of the Institute for
Theoretical Physics (ITP) in Santa Barbara, CA, where,
during the last month of the Cosmological Phase Transitions research
program, this work began. This work was supported in part by
a National Science Foundation Grant n. PHY89-04035 at ITP and by
Conselho Nacional
de Desenvolvimento Cient\'{\i}fico e Tecnol\'ogico - CNPq (Brazil).

\end{document}